\documentclass[prl,floatfix,reprint,amsmath,amssymb,aps]{revtex4-2}

\usepackage{graphicx}
\usepackage{stfloats}
\usepackage{float}
\usepackage{dcolumn}
\usepackage{bm}
\usepackage{amsmath}
\usepackage{amssymb}

\usepackage[colorlinks=true, allcolors=blue]{hyperref}

\begin{document}

\preprint{APS/123-QED}

\title{Drift-Induced Nonreciprocal Hyperbolic Polaritons in Graphene/\texorpdfstring{$\alpha$-MoO$_3$}{alpha-MoO3} Heterostructures}

\author{Rajveer Fandan}
\email{rajveer.fandan@upm.es}
\affiliation{Instituto de Sistemas Optoelectrónicos y Microtecnología, Universidad Politécnica de Madrid, Av. Complutense 30, Madrid 28040, Spain}
\affiliation{Departamento de Ingeniería Electrónica, E.T.S.I de Telecomunicación, Universidad Politécnica de Madrid, Av. Complutense 30, Madrid 28040, Spain}

\author{Jorge Pedrós}
\email{j.pedros@upm.es}
\affiliation{Instituto de Sistemas Optoelectrónicos y Microtecnología, Universidad Politécnica de Madrid, Av. Complutense 30, Madrid 28040, Spain}
\affiliation{Departamento de Ingeniería Electrónica, E.T.S.I de Telecomunicación, Universidad Politécnica de Madrid, Av. Complutense 30, Madrid 28040, Spain}

\begin{abstract}
Achieving optical isolation requires breaking symmetry between forward- and backward-propagating light, a long-standing challenge at the nanoscale in the absence of magnetic fields. Here we theoretically demonstrate electrically tunable nonreciprocal phonon–plasmon polaritons in a graphene/$\alpha$-MoO$_3$/SiC heterostructure operating in the mid-infrared. A dc current in graphene induces a wavevector-dependent Doppler shift that breaks reciprocity and generates strong directional asymmetry in hybrid plasmon–phonon propagation. In the reciprocal regime, hybridization between graphene plasmons and hyperbolic phonon polaritons in $\alpha$-MoO$_3$, further shaped by the SiC substrate, enables gate-controlled transitions of isofrequency contours, including canalization along orthogonal crystal axes. At drift velocities of $5\%$ of the Fermi velocity, the system exhibits pronounced momentum-dependent nonreciprocity with contrast reaching $\sim 0.3$, while directions orthogonal to the drift remain unaffected due to symmetry-imposed constraints. Real-space calculations confirm that this momentum-space asymmetry translates into directional near-field intensity modulation. These results establish current-biased van der Waals heterostructures as a platform for electrically tunable, magnet-free nonreciprocal nanophotonics in the mid-infrared.
\end{abstract}

\maketitle

\section{\label{sec:intro}Introduction}

Breaking reciprocity in light propagation is essential for optical isolation and circulators, but conventional approaches rely on bulky magneto-optical materials that are difficult to integrate at the nanoscale~\cite{Jalas2013,bi2011chip}. This has motivated compact, magnet-free schemes based on spatiotemporal modulation~\cite{yu2009complete,sounas2017non}, optomechanics~\cite{ruesink2016nonreciprocity}, and dc-biased electron systems~\cite{hassani2022drifting,Mazor2019}. Among these, drift-induced nonreciprocity is particularly attractive, as a steady current in a conductor produces a wavevector-dependent Doppler shift $\mathbf{q}\cdot\mathbf{v}_{\mathrm d}$, where $\mathbf{v}_{\mathrm d}$ is the carrier drift velocity associated with the applied dc current, breaking symmetry between opposite propagation directions without external fields or moving parts~\cite{ActiveGraphene2021,Morgado2018,CorreasSerrano2019,dutta2024large,xu2025electrically}.

Graphene is an ideal platform for this effect due to its high mobility, tunable Fermi energy, and Dirac dispersion~\cite{RevModPhys.81.109,RevModPhys.83.407}. Experimentally accessible drift velocities reach a significant fraction of the Fermi velocity~\cite{Shin2018}, enabling asymmetric plasmon propagation. In isotropic environments such as SiO$_2$ or h-BN, this response remains aligned with the current direction~\cite{zhao2021efficient,dong2021fizeau}. A qualitatively different regime emerges when graphene is coupled to anisotropic polaritonic media, where the interplay between carrier drift and directional phonon polaritons remains unexplored.

The biaxial van der Waals crystal $\alpha$-MoO$_3$ provides natural in-plane hyperbolicity, where opposite signs of the in-plane dielectric permittivity components support highly directional phonon polaritons within its Reststrahlen bands~\cite{dong2020broad,ma2018plane}. When integrated with graphene, these modes hybridize into tunable phonon–plasmon polaritons with gate-dependent isofrequency contours (IFCs)~\cite{alvarez2022active,zhou2023gate,hu2022doping}. The SiC substrate further reshapes the electrostatic environment through its phononic response, enhancing confinement and enabling canalized propagation without nanostructuring~\cite{garcia2025modulation,duan2025canalization}. Together, these materials form a platform where electrostatic gating, anisotropy, and carrier drift jointly control polariton propagation.

Here, we study drift-induced nonreciprocity in graphene/$\alpha$-MoO$_3$/SiC heterostructures using a drift-corrected random phase approximation (RPA) framework and numerical simulations. We show that anisotropic hyperbolic dispersion fundamentally reshapes drift-induced symmetry breaking: while the zero-contrast direction remains pinned perpendicular to the current independent of material anisotropy, the magnitude of nonreciprocity is strongly governed by the underlying IFC geometry. Gate-controlled transitions between elliptical, hyperbolic, and canalized regimes enable efficient tuning of this response, with maximum contrast reaching $\sim0.3$ at realistic drift velocities ($5\%$ of the Fermi velocity). Momentum- and real-space analyses reveal a direct link between anisotropic dispersion, drift-induced Doppler shifts, and observable near-field asymmetry.

\section{\label{sec:level2}Coulomb interaction, dielectric screening, and drift-modified loss function}

We consider a graphene/$\alpha$-MoO$_3$/SiC heterostructure in which the SiC substrate occupies the lower half-space and the $\alpha$-MoO$_3$ layer forms a biaxial van der Waals slab of thickness $d$. The crystalline axes [100], [001], and [010] define the in-plane $(x,y)$ and out-of-plane $(z)$ directions, yielding a frequency-dependent dielectric tensor $\boldsymbol{\varepsilon}(\omega)$ (see Supporting Information Section 1).

Graphene, a two-dimensional semimetal with massless Dirac fermions, exhibits a linear energy dispersion near the Dirac point. The dielectric environment dynamically screens the Coulomb interaction between graphene charge carriers, giving rise to collective plasmon excitations~\cite{yan2013damping,hwang2010plasmon}. The surrounding media further modify the Coulomb interaction and therefore the plasmon dispersion~\cite{tomadin2015accessing,fandan2018acoustically}. Within the RPA framework, the dielectric screening function is

\begin{equation}
    \varepsilon_{\mathrm{RPA}}(\mathbf{q},\omega)
    = 1 - V(\mathbf{q},\omega)\,\Pi(\mathbf{q},\omega),
    \label{eq:epsilon_rpa}
\end{equation}

where $\mathbf{q} = (q_x, q_y)$ is the in-plane wavevector, $\Pi(\mathbf{q},\omega)$ is the irreducible graphene polarizability, and $V(\mathbf{q},\omega)$ is the effective Coulomb kernel describing the electrostatic response of the vacuum/MoO$_3$/SiC multilayer structure. Solving the electrostatic boundary-value problem for a surface charge at the graphene plane in the presence of the anisotropic MoO$_3$ slab and SiC half-space yields~\cite{fandan2019effect,tomadin2015accessing}

\begin{equation}
  V(\mathbf{q},\omega)
  = v_c(q,\omega)\,
    \frac{\varepsilon_z q\kappa
        + \varepsilon_{\mathrm{SiC}} q^{2}\tanh(\kappa d)}
         {\varepsilon_z q\kappa
        + \frac{\varepsilon_{\mathrm{SiC}} q^{2}
               + \varepsilon_z^{2}\kappa^{2}}
              {\varepsilon_{\mathrm{SiC}}+1}
          \tanh(\kappa d)},
    \label{eq:V_kernel}
\end{equation}

where $v_c(q,\omega) = e^{2}/[2\varepsilon_0 q(1+\varepsilon_{\mathrm{SiC}})]$ is the substrate-screened Coulomb interaction in the absence of the slab and $\kappa(\mathbf{q},\omega) = \sqrt{(\varepsilon_x q_x^2+\varepsilon_y q_y^2)/\varepsilon_z}$ is the anisotropic electrostatic decay constant inside the slab (see Supporting Information Section 2). Since all $\mathbf{q}$-dependence enters through $q=|\mathbf{q}|$ and through $\kappa^2(\mathbf{q},\omega)\propto \varepsilon_x q_x^2+\varepsilon_y q_y^2$, the kernel is invariant under momentum inversion,
$V(-\mathbf{q}, \omega)=V(\mathbf{q}, \omega)$. Because the equilibrium graphene polarizability is isotropic, the anisotropy of $\varepsilon_{\mathrm{RPA}}$ originates entirely from $V(\mathbf{q},\omega)$. We neglect local-field (exchange--correlation) corrections, which become relevant only at large wavevectors and low carrier densities~\cite{fandan2019effect}.

To include carrier drift, we introduce a finite average drift velocity $\mathbf{v}_{\mathrm{d}}$ induced by a dc in-plane electric field. Drift breaks time-reversal symmetry and modifies the carrier distribution to $f_{\mathbf{v}_{\mathrm{d}}}(\varepsilon_{\mathbf{k}}) = f_0[\varepsilon_{\mathbf{k}} - \hbar \mathbf{k}\cdot\mathbf{v}_{\mathrm{d}}]$~\cite{Svintsov2018,Morgado2018,CorreasSerrano2019}.
We express $\mathbf{v}_{\mathrm{d}}$ as a fraction of the graphene Fermi velocity $v_F$:
\begin{equation}
    \mathbf{v}_{\mathrm{d}} = \beta v_F\,\hat{\mathbf{v}}_{\mathrm{d}},
\end{equation}
where $0\leq\beta < 1$ is a dimensionless drift parameter.
For the parameter range considered here, with Fermi energy $E_F = 0.1~\mathrm{eV}$ (corresponding to a carrier density $n \approx 7 \times 10^{11}~\mathrm{cm}^{-2}$) and $v_F = 10^6~\mathrm{m/s}$, the choice $\beta = 0.05$ gives $v_{\mathrm{d}} = 5 \times 10^{4}~\mathrm{m/s}$. This value corresponds to current densities of approximately $0.2~\mathrm{mA/\mu m}$, placing our calculations within the experimentally accessible regime of dc-biased graphene plasmonics~\cite{Shin2018,zhao2021efficient,dong2021fizeau}. Experiments have reported current densities up to $\sim 0.7~\mathrm{mA/\mu m}$, corresponding to drift velocities as high as $v_{\mathrm{d}}/v_F \approx 0.17$~\cite{dong2021fizeau}. Our choice of $\beta = 0.05$ is therefore conservative and remains well below the thermal and nonlinear regimes associated with extreme driving.

Within the RPA framework, drift introduces a Doppler-like frequency shift in the polarizability~\cite{wunsch2006dynamical,sabbaghi2015drift}:
\begin{equation}
\begin{split}
\Pi_{\mathrm{drift}}(\mathbf{q},\omega)
= \Pi_0\!\left(\mathbf{q},\,\omega - \delta_{\mathbf{q}}\right), \\
\delta_{\mathbf{q}} \equiv \mathbf{q}\cdot\mathbf{v}_{\mathrm{d}}
= \beta v_F q\cos(\varphi_{\mathbf{q}} - \theta_d),
\end{split}
\label{eq:drift_polarizability_beta}
\end{equation}
where $\Pi_0$ is the equilibrium polarizability, $\delta_{\mathbf{q}}$ is the Doppler detuning, $\varphi_{\mathbf{q}}$ is the polar angle of $\mathbf{q}$, and $\theta_d$ is the drift-current angle. We use Eq.~(\ref{eq:drift_polarizability_beta}) directly in the numerical evaluation of the density response. In the equivalent conductivity formulation, a laboratory-frame prefactor appears in the drifted Drude conductivity to ensure charge conservation; however, when the density response is reconstructed from the conductivity through the continuity relation, this prefactor cancels out, leaving the Doppler-shifted form of Eq.~(\ref{eq:drift_polarizability_beta}) for $\Pi_{\mathrm{drift}}$ (see Supporting Information Section 2). The corresponding drift-modified dielectric screening function is
\begin{equation}
    \varepsilon_{\mathrm{RPA}}^{\mathrm{drift}}(\mathbf{q},\omega)
    = 1 - V(\mathbf{q},\omega)\,\Pi_0\!\left(\mathbf{q},\,\omega - \delta_{\mathbf{q}}\right).
    \label{eq:epsilon_drift_beta}
\end{equation}

Because the Doppler term is odd under momentum inversion \cite{caloz2018electromagnetic}, $\delta_{-\mathbf{q}}=-\delta_{\mathbf{q}}$, whereas the interaction kernel is even, $V(-\mathbf{q},\omega)=V(\mathbf{q},\omega)$, reversing $\mathbf{q}$ at fixed drift is equivalent to reversing the drift velocity at fixed $\mathbf{q}$. The response therefore remains reciprocal whenever $\delta_{\mathbf{q}}=0$, i.e., for $\mathbf{q}\perp\mathbf{v}_{\mathrm{d}}$. This condition defines a direction determined solely by the drift axis and independent of substrate anisotropy. We demonstrate this \emph{axis-pinning theorem} in the Supporting Information Section 3.

We quantify nonreciprocity through the drift-modified loss function

\begin{equation}
    L(\mathbf{q}, \omega) = -\mathrm{Im}\!\left[\frac{1}{\varepsilon_{\mathrm{RPA}}^{\mathrm{drift}}(\mathbf{q},\omega)}\right],
    \label{eq:loss_function_beta}
\end{equation}

and the dimensionless nonreciprocal contrast

\begin{equation}
    C(\mathbf{q}, \omega) =
    \frac{L(+\mathbf{q},\omega) - L(-\mathbf{q},\omega)}
         {L(+\mathbf{q},\omega) + L(-\mathbf{q},\omega)},
    \label{eq:contrast_function}
\end{equation}

where $C=0$ corresponds to reciprocal behavior and $|C|\to 1$ to complete directional isolation. The condition $\delta_{\mathbf{q}}=0$ therefore maps directly onto $C(\mathbf{q},\omega)=0$, identifying the zero-contrast direction perpendicular to the drift current. For weak drift, a first-order expansion of the drift-modified response (see Supporting Information, Section 3) yields
\begin{equation}
C(\mathbf q,\omega)\simeq
\left(\frac{v_d}{v_{\mathrm{ph}}}\right)
\cos(\varphi_{\mathbf q}-\theta_d)\,
\mathcal F(\mathbf q,\omega;E_F,d),
\label{eq:contrast_approximate}
\end{equation}
where $v_{\mathrm{ph}}=\omega/q$ is the phase velocity, and $\mathcal F$ is a reciprocal enhancement factor determined by the equilibrium polaritonic response of the heterostructure. This expression makes explicit that the nodal direction is fixed solely by the Doppler projection factor, whereas the magnitude of the contrast is governed by the reciprocal polaritonic environment and can therefore exhibit a strong dependence on the anisotropic IFC geometry.


\section{\label{sec:level3}Results and Discussion}

\subsection{Reciprocal polariton landscape and gate-tunable topological transitions}

\begin{figure*}[!t]
    \centering
    \includegraphics[width=\textwidth]{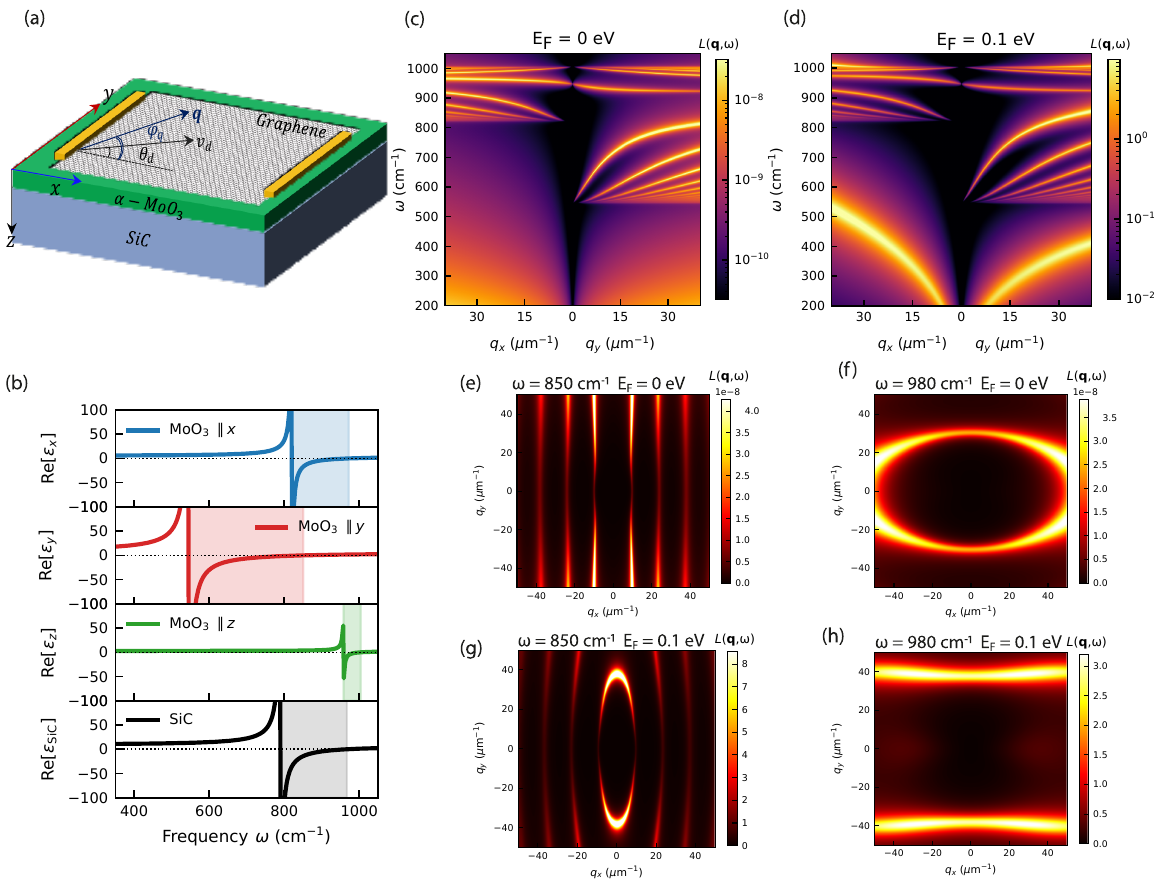}
    \caption{\textbf{Graphene/$\alpha$-MoO$_3$/SiC heterostructure and tunable polaritonic dispersion.}
\textbf{(a)} Schematic of the graphene/$\alpha$-MoO$_3$/SiC device geometry. 
\textbf{(b)} Real parts of the relative permittivities of $\alpha$-MoO$_3$ and SiC versus frequency. The shaded regions denote the Reststrahlen bands bounded by the transverse optical (TO) and longitudinal optical (LO) phonon frequencies. 
\textbf{(c)} Polariton dispersion of the graphene/$\alpha$-MoO$_3$ ($d = 100~\mathrm{nm}$)/SiC structure for $E_F = 0~\mathrm{eV}$, calculated from the loss function with $\beta = 0$ [Eq.~(\ref{eq:loss_function_beta})]. Without graphene plasmons, the spectrum is dominated by the phonon-polariton modes of $\alpha$-MoO$_3$, with multiple hyperbolic branches along both the $x$- and $y$-directions within the Reststrahlen bands. 
\textbf{(d)} Same as in \textbf{(c)}, but for $E_F = 0.1~\mathrm{eV}$. Finite graphene doping hybridizes graphene plasmons with the $\alpha$-MoO$_3$ phonon-polariton modes, strongly modifying the dispersion and enabling electrical tunability through the graphene Fermi energy. 
\textbf{(e,f)} IFCs at $\omega = 850~\mathrm{cm}^{-1}$ and $\omega = 980~\mathrm{cm}^{-1}$, respectively, for $E_F = 0~\mathrm{eV}$. At $\omega = 850~\mathrm{cm}^{-1}$, the IFC exhibits canalized propagation along the $x$-direction, whereas at $\omega = 980~\mathrm{cm}^{-1}$ the IFC is elliptical. 
\textbf{(g,h)} Same as in \textbf{(e,f)}, but for $E_F = 0.1~\mathrm{eV}$. Finite graphene doping strongly modifies the IFC topology: the contour at $850~\mathrm{cm}^{-1}$ becomes elliptical, while the contour at $\omega = 980~\mathrm{cm}^{-1}$ evolves into open hyperbolic arcs with canalization along the $y$-direction.
}
    \label{fig:placeholder1}
\end{figure*}

We first examine the polaritonic response of the graphene/$\alpha$-MoO$_3$/SiC heterostructure in the absence of carrier drift. Figure~\ref{fig:placeholder1}(a) shows the device geometry, consisting of a graphene monolayer on a thin $\alpha$-MoO$_3$ slab ($d=100~\mathrm{nm}$) supported by a SiC substrate. Figure~\ref{fig:placeholder1}(b) shows the real parts of the principal permittivities. The Reststrahlen bands of $\alpha$-MoO$_3$ lie between the transverse optical (TO) and longitudinal optical (LO) phonon frequencies along each crystallographic axis, while the SiC substrate introduces an additional phonon resonance whose negative-permittivity region overlaps with the $y$-direction Reststrahlen band of $\alpha$-MoO$_3$. This overlap strongly influences the Coulomb kernel $V(\mathbf{q},\omega)$ [Eq.~(\ref{eq:V_kernel})]. When $\varepsilon_{\mathrm{SiC}}$ becomes large and negative, the substrate acts as an electrostatic mirror that enhances field confinement and promotes canalization without requiring patterning or twisting of the $\alpha$-MoO$_3$ crystal\cite{garcia2025modulation}.

For the undoped case ($E_F = 0~\mathrm{eV}$), the polariton dispersion is dominated by the intrinsic phonon-polariton modes of $\alpha$-MoO$_3$, exhibiting multiple hyperbolic branches along both the $x$- and $y$-directions within the Reststrahlen bands [Fig.~\ref{fig:placeholder1}(c)]. Finite graphene doping ($E_F = 0.1~\mathrm{eV}$) introduces a plasmonic contribution to the polarizability $\Pi_0(\mathbf{q},\omega)$ that hybridizes with the phonon polaritons, strongly modifying the dispersion and opening anticrossing gaps characteristic of phonon--plasmon coupling [Fig.~\ref{fig:placeholder1}(d)]. This hybridization is controlled by the Drude weight $D = e^2E_F/(\pi\hbar^2)$, which determines the graphene contribution to $\varepsilon_{\mathrm{RPA}}$ and can be continuously tuned through electrostatic gating.

IFCs provide complementary insight into the directional propagation of the polaritons. At $\omega = 850~\mathrm{cm}^{-1}$ and $E_F = 0~\mathrm{eV}$, the IFC exhibits a nearly flat canalized geometry with propagation concentrated along the $x$-direction [Fig.~\ref{fig:placeholder1}(e)], whereas at $\omega = 980~\mathrm{cm}^{-1}$ it forms a closed elliptical contour [Fig.~\ref{fig:placeholder1}(f)]. Finite graphene doping qualitatively reshapes the IFC topology: the contour at $850~\mathrm{cm}^{-1}$ closes into an ellipse [Fig.~\ref{fig:placeholder1}(g)], while the contour at $980~\mathrm{cm}^{-1}$ opens into hyperbolic arcs with canalization along the $y$-direction [Fig.~\ref{fig:placeholder1}(h)]. These gate-induced topological transitions---from open (hyperbolic) to closed (elliptical) contours and vice versa---arise because doping modifies the graphene plasmon contribution to the hybrid modes, effectively shifting the frequency range over which the condition $\mathrm{Re}[\varepsilon_x]\mathrm{Re}[\varepsilon_y] < 0$ is satisfied in the dressed system. As shown below, the high-momentum and strongly directional channels associated with canalization provide an effective route for enhancing the nonreciprocal response.

\subsection{Drift-induced nonreciprocal contrast in momentum space}

\begin{figure*}[!t]
    \centering
    \includegraphics[width=\textwidth]{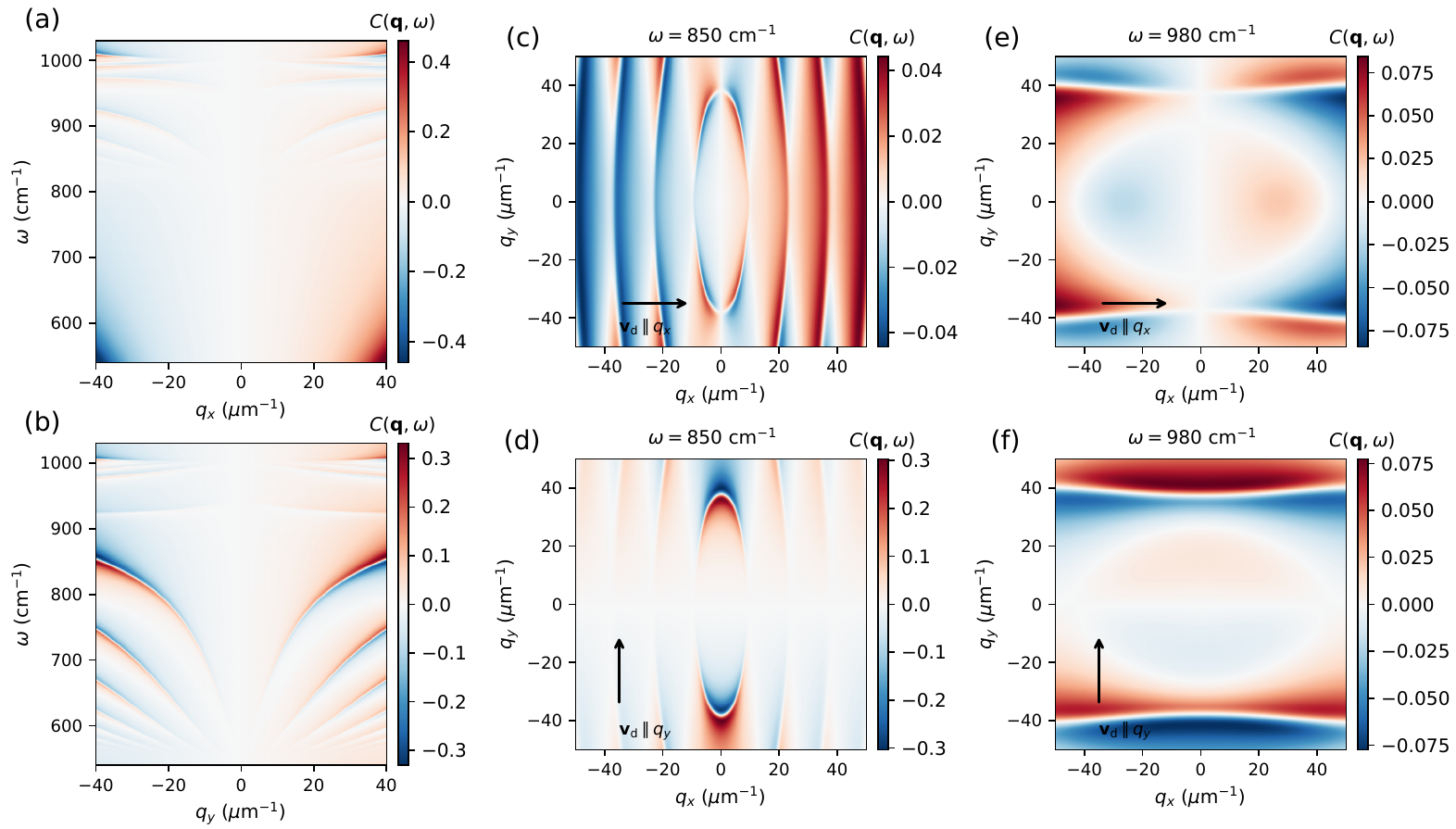}
    \caption{\textbf{Drift-induced nonreciprocal contrast in the polariton dispersion and IFCs.}
\textbf{(a,b)} Momentum-space contrast induced by a carrier drift with $\beta = 0.05$, computed from the loss-function asymmetry [Eq.~(\ref{eq:loss_function_beta})] along the $q_x$ and $q_y$ directions, respectively. In \textbf{(a)}, the drift is aligned with $q_x$, while in \textbf{(b)} it is aligned with $q_y$. In both cases, the drift breaks the $\mathbf{q}\leftrightarrow -\mathbf{q}$ symmetry of the loss function, leading to nonreciprocal polariton dispersion. When the drift is orthogonal to the propagation direction, the dispersion remains nearly unchanged because $\mathbf{q}\cdot\mathbf{v}_{\mathrm{d}}=0$.
\textbf{(c,d)} Contrast maps of the IFCs at $\omega = 850~\mathrm{cm}^{-1}$ for drift along the $x$ and $y$ directions, respectively. The IFCs show clear asymmetry between the $+\mathbf{q}$ and $-\mathbf{q}$ branches along the drift direction, while remaining approximately symmetric in the transverse direction.
\textbf{(e,f)} Same as in \textbf{(c,d)}, but for $\omega = 980~\mathrm{cm}^{-1}$. Parameters: $E_F = 0.1~\mathrm{eV}$ and $d = 100~\mathrm{nm}$.
}
    \label{fig:placeholder2}
\end{figure*}

We now introduce a finite carrier drift with $\beta = 0.05$ and examine its effect on the polariton spectrum. Figures~\ref{fig:placeholder2}(a) and~\ref{fig:placeholder2}(b) show the momentum-resolved nonreciprocal contrast $C(\mathbf{q},\omega)$ [Eq.~(\ref{eq:contrast_function})] evaluated along the $q_x$ and $q_y$ directions, respectively, with the drift aligned parallel to the propagation axis in each case. In practice, we compute the contrast by comparing opposite drift directions at fixed $\mathbf{q}$, which is equivalent to the $\pm\mathbf{q}$ definition in Eq.~(\ref{eq:contrast_function}) within the present drift-shift model. The $\mathbf{q}\leftrightarrow -\mathbf{q}$ symmetry of the $L(\mathbf{q}, \omega)$ is broken along the drift direction, producing a clear frequency splitting between forward ($+\mathbf{q}$) and backward ($-\mathbf{q}$) polaritons. In contrast, when the drift is orthogonal to the propagation direction, the contrast vanishes within numerical precision. This directly reflects the axis-pinning theorem: the Doppler detuning $\delta_{\mathbf{q}} = \mathbf{q}\cdot\mathbf{v}_{\mathrm{d}}$ vanishes for $\mathbf{q}\perp\mathbf{v}_{\mathrm{d}}$, preventing any odd-in-$\mathbf{q}$ response regardless of substrate anisotropy.

We further analyze the nonreciprocal response using IFC contrast maps. At $\omega = 850~\mathrm{cm}^{-1}$, where the doped IFC is elliptical [Fig.~\ref{fig:placeholder1}(g)], the contrast maps for drift along the $x$ and $y$ directions [Figs.~\ref{fig:placeholder2}(c,d)] show a dipolar pattern with positive $C(\mathbf{q}, \omega)$ (enhanced spectral weight) downstream and negative $C(\mathbf{q}, \omega)$ upstream, and nodal lines perpendicular to the drift direction. At $\omega = 980~\mathrm{cm}^{-1}$, where the doped IFC is open and canalized along $y$ [Fig.~\ref{fig:placeholder1}(h)], the nonreciprocal contrast becomes significantly stronger [Figs.~\ref{fig:placeholder2}(e,f)], reaching values of approximately ${\sim}0.3$ in the most strongly confined regions of the IFC.

Although the axis-pinning theorem fixes the direction of zero contrast to be perpendicular to the drift, the magnitude of $C(\mathbf{q}, \omega)$ along a given contour is governed jointly by the Doppler projection factor $\cos(\varphi_{\mathbf{q}} - \theta_d)$ and the direction-dependent confinement encoded in the IFC geometry. This interplay produces the lobed patterns in Figs.~\ref{fig:placeholder2}(c--f), where the contrast is strongest along IFC regions with large $q$ components aligned with the drift direction and weakest (aside from the nodal line) where either the $q$ magnitude is small or the projection factor vanishes. Substrate anisotropy therefore modulates the amplitude of the nonreciprocal response without altering its symmetry axis, reflecting a clear separation of roles that holds within the drift-shift framework.

Consistent with Eq.~(\ref{eq:contrast_approximate}), the nonreciprocal contrast can be tuned independently through the drift strength, graphene Fermi energy, and $\alpha$-MoO$_3$ thickness. Increasing the drift velocity enhances the prefactor $v_d/v_{\mathrm{ph}}$ and therefore the overall contrast magnitude, while variations in $E_F$ and $d$ modify the reciprocal enhancement factor $\mathcal F$ through changes in the hybrid polariton dispersion. A systematic parameter sweep demonstrating the dependence of $C(\mathbf q,\omega)$ on $\beta$, $E_F$, $d$, and drift orientation ($\theta_d$) is provided in Supporting Information Section~5. We further examine the influence of graphene electron damping ($\gamma_e$) in Supporting Information Section~5. While the contrast remains relatively insensitive to damping for propagation along the $q_x$ direction, a stronger suppression is observed for modes propagating along $q_y$. This anisotropic damping dependence reflects the hybrid plasmon--phonon character of the polaritons: modes exhibiting larger nonreciprocal contrast possess a greater graphene-plasmon contribution and are therefore more susceptible to electronic losses, whereas more phonon-dominated modes remain comparatively robust. We note that the maximum contrast values ($\sim 0.3$) remain stable for realistic graphene damping rates reported experimentally, while only moderate quantitative reductions are observed upon increasing $\gamma_e$ (see Supporting Information Section~5).

\begin{figure}[t]
    \centering
    \includegraphics[width=\columnwidth]{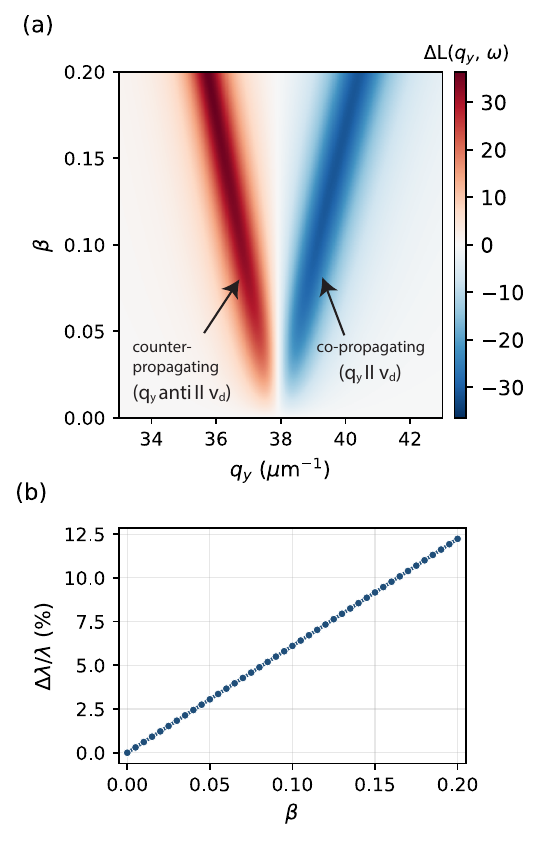}
    \caption{\textbf{Drift-induced nonreciprocity and wavelength asymmetry of graphene--MoO$_3$ hybrid polaritons.} \textbf{a,} Density map of the differential loss function
    $\Delta L(q_y,\omega)=L(+q_y,\omega)-L(-q_y,\omega)$
    at $\omega=850~\mathrm{cm}^{-1}$ for a graphene/MoO$_3$ (100 nm)/SiC heterostructure with $E_F=0.1~\mathrm{eV}$. The electron drift velocity is along $+y$ and parameterized by $\beta=v_d/v_F$. At $\beta=0$, the system is reciprocal and $\Delta L=0$. Finite drift breaks $q\rightarrow -q$ symmetry through a Doppler shift of the graphene response, producing momentum splitting between counter-propagating hybrid plasmon--phonon polaritons. The co-propagating mode ($q_y\parallel\beta$) shifts to larger wavevectors (shorter wavelength), while the counter-propagating mode ($q_y\,\mathrm{anti}\parallel\,\beta$) shifts to smaller wavevectors (longer wavelength). \textbf{b,} Relative wavelength asymmetry,
    $\Delta\lambda/\lambda$,
    obtained from the peak positions of the loss function. The drift-induced momentum splitting leads to wavelength compression for the co-propagating mode and wavelength expansion for the counter-propagating mode, resulting in a monotonic increase of asymmetry with $\beta$.}
    \label{fig:asymmetry}
\end{figure}

An alternative and experimentally intuitive measure of the drift-induced nonreciprocity is obtained by directly tracking the displacement of the polariton resonances in momentum space. To quantify the directional imbalance in the plasmonic response, we define the differential loss function as $\Delta L(\mathbf{q},\omega)=L(\mathbf{q},\omega)-L(-\mathbf{q},\omega)$, which compares the loss associated with opposite propagation directions at the same frequency. Figure~\ref{fig:asymmetry}(a) shows $\Delta L(q_y,\omega)$ evaluated at a fixed frequency of $850~\mathrm{cm}^{-1}$ as a function of the drift strength, where the wavevector is taken along the $y$ direction. In the reciprocal limit ($\beta=0$), the forward- and backward-propagating modes are degenerate, resulting in $\Delta L=0$ throughout the spectrum. Finite drift lifts this degeneracy through the Doppler shift $\delta_{\mathbf q}=\mathbf q\cdot\mathbf v_{\mathrm d}$, producing a progressive momentum splitting between the two counter-propagating polariton branches. Positive values of $\Delta L$ indicate that the loss intensity of the $+q_y$ branch exceeds that of the $-q_y$ branch, whereas negative values indicate the opposite. As the drift velocity increases, the co-propagating mode ($q_y\parallel\mathbf{v}_{\mathrm d}$) shifts toward larger wavevectors, corresponding to stronger confinement and shorter polariton wavelengths, while the counter-propagating mode ($q_y$ antiparallel to $\mathbf{v}_{\mathrm d}$) shifts toward smaller wavevectors and longer wavelengths. The appearance of distinct positive and negative ridges in the differential loss map therefore reflects both the drift-induced redistribution of spectral weight and the increasing separation of the two resonances in momentum space.

The resulting nonreciprocal wavelength splitting is quantified in Fig.~\ref{fig:asymmetry}(b), which plots the relative wavelength asymmetry extracted from the wavevectors corresponding to the maxima of the loss function for the co-propagating and counter-propagating polariton branches. Specifically, the asymmetry is defined as $\Delta\lambda/\lambda=(\lambda_{+}-\lambda_{-})/\left[(\lambda_{+}+\lambda_{-})/2\right]$, where $\lambda_{+}$ and $\lambda_{-}$ denote the polariton wavelengths associated with propagation parallel and antiparallel to the drift velocity, respectively. The quantity shown therefore measures the relative separation between the two nonreciprocal branches, rather than the shift of either branch with respect to the equilibrium ($\beta=0$) dispersion. Because the co-propagating mode is displaced to larger $q$, its associated polariton wavelength decreases, whereas the wavelength of the counter-propagating mode increases. Consequently, the wavelength asymmetry grows monotonically with drift velocity, reflecting the increasing separation between the two branches. For sufficiently small drift velocities, where the Doppler-induced shifts are approximately symmetric about the equilibrium dispersion, the average wavelength satisfies $(\lambda_{+}+\lambda_{-})/2 \approx \lambda_{0}$, where $\lambda_{0}$ is the equilibrium polariton wavelength. In this limit, the relative wavelength shift of either individual branch with respect to equilibrium, $(\lambda_{\pm}-\lambda_{0})/\lambda_{0}$, is approximately one-half of the asymmetry plotted in Fig.~\ref{fig:asymmetry}(b). Unlike the contrast function $C(\mathbf q,\omega)$, which measures differences in spectral weight, the wavelength asymmetry directly characterizes the nonreciprocal modification of the polariton dispersion and therefore provides a complementary metric for quantifying electrically-controlled directional propagation in graphene/$\alpha$-MoO$_3$/SiC heterostructures.

\subsection{Real-space signatures of nonreciprocal polariton propagation}

To connect the momentum-space description with experimentally accessible near-field observables, we reconstruct the real-space screened potential $\phi_{\beta}(\mathbf{r})$ via inverse Fourier transform of the momentum-space response kernel (see Supporting Information Section~4) \cite{zhang2022strong,schwartz2021substrate}. The excitation is modeled as a localized out-of-plane point dipole source positioned $z_h = 10~\mathrm{nm}$ above the graphene layer, with $E_F = 0.1~\mathrm{eV}$ and drift parameter $\beta = 0.05$.

\begin{figure*}[!t]
    \centering
    \includegraphics[width=\textwidth]{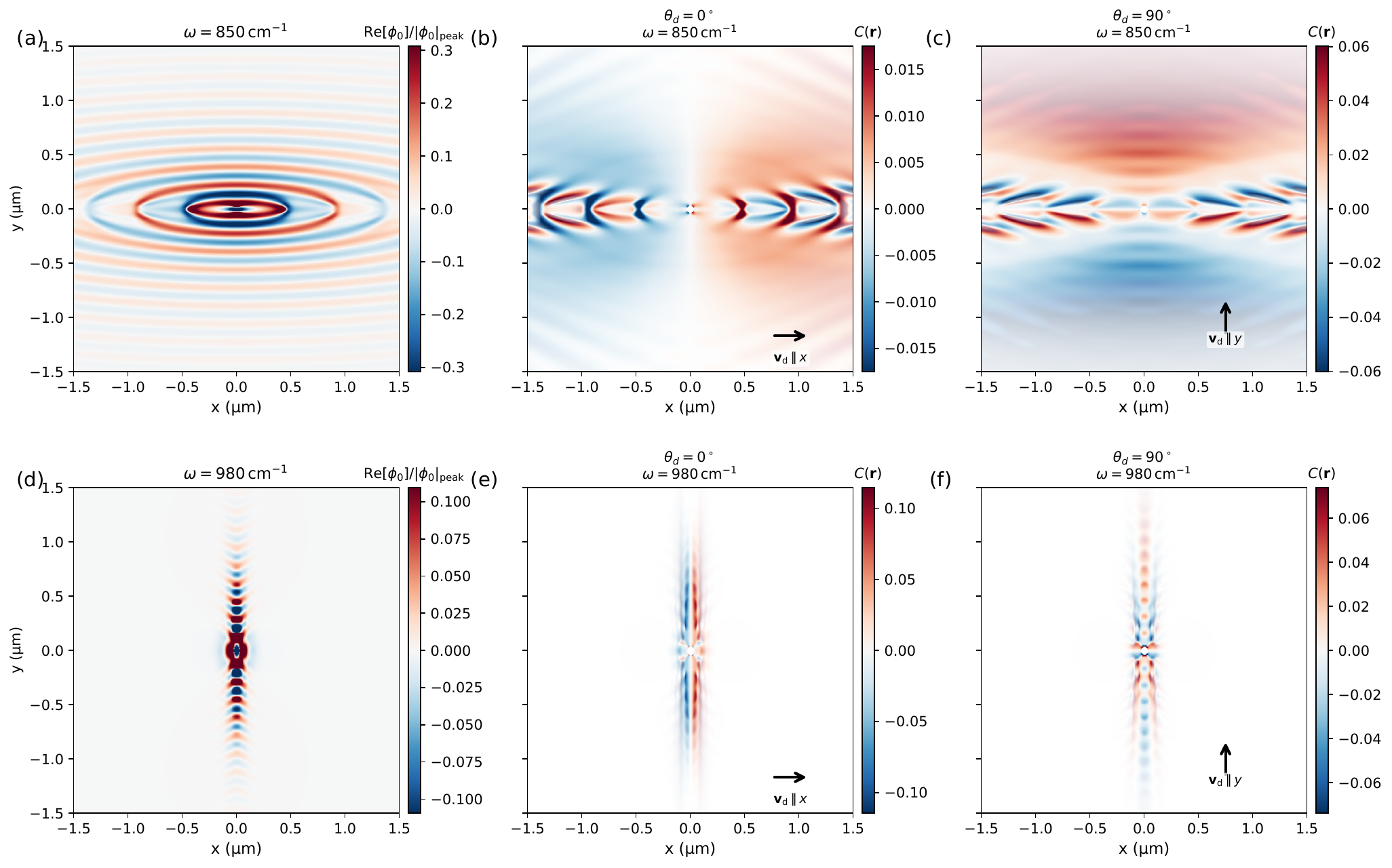}
    \caption{\textbf{Frequency- and direction-resolved real-space signatures of nonreciprocal polariton propagation under carrier drift.}
\textbf{(a,d)} Real part of the equilibrium potential, $\mathrm{Re}[\phi_0]/|\phi_0|_{\mathrm{peak}}$, in the absence of drift ($\beta = 0$), at excitation frequencies $\omega = 850~\mathrm{cm}^{-1}$ (top row) and $\omega = 980~\mathrm{cm}^{-1}$ (bottom row), showing reciprocal polariton fringes launched by a point source.
\textbf{(b,e)} Spatial contrast map $C(\mathbf{r})$ [Eq.~(\ref{eq:contrast_real_space})] for carrier drift along the $x$-direction ($\theta_d = 0^\circ$).
\textbf{(c,f)} Same as \textbf{(b,e)}, but for drift along the $y$-direction ($\theta_d = 90^\circ$).
The contrast isolates the non-reciprocal component of the response by comparing opposite drift configurations $\pm\beta$, with red (blue) regions indicating enhanced (suppressed) near-field amplitude under reversal of the drift direction.
The arrows denote the carrier drift directions.
System parameters: $z_h = 10~\mathrm{nm}$, $E_F = 0.1~\mathrm{eV}$, $d = 100~\mathrm{nm}$, and $\beta = 0.05$.}
    \label{fig:placeholder3}
\end{figure*}

Figure~\ref{fig:placeholder3} summarizes the resulting near-field response at two representative frequencies, $\omega = 850~\mathrm{cm}^{-1}$ and $\omega = 980~\mathrm{cm}^{-1}$. The equilibrium panels [Figs.~\ref{fig:placeholder3}(a) and \ref{fig:placeholder3}(d)] show $\mathrm{Re}[\phi_0(\mathbf{r})]/|\phi_0|_{\mathrm{peak}}$, obtained for $\beta = 0$. In both cases, the field pattern is strongly anisotropic, reflecting directional polariton propagation imposed by the $\alpha$-MoO$_3$/SiC environment. The two frequencies also produce distinct beam geometries and fringe structures, indicating different redistributions of momentum-space spectral weight across the corresponding IFCs.

To isolate nonreciprocity, we define the real-space contrast which directly measures the change in local field intensity under reversal of the drift direction
\begin{equation}
C(\mathbf{r})=\frac{|\phi_{+\beta}(\mathbf{r})|-|\phi_{-\beta}(\mathbf{r})|}{|\phi_{+\beta}(\mathbf{r})|+|\phi_{-\beta}(\mathbf{r})|},
\label{eq:contrast_real_space}
\end{equation}
where $\phi_{\pm\beta}$ correspond to opposite drift directions. This quantity filters out reciprocal propagation channels and retains only the odd-in-$\mathbf{q}$ response associated with drift-induced Doppler shifts. Physically, $C(\mathbf{r})$ arises from the interference between drift-shifted momentum components of the IFC, so that real-space asymmetry directly encodes the anisotropic redistribution of spectral weight in momentum space.

For drift along $x$, the contrast [Figs.~\ref{fig:placeholder3}(b,e)] shows a clear downstream–upstream asymmetry localized along dominant propagation channels, with magnitude set by the overlap between Doppler-shifted modes and the anisotropic IFC. Rotating the drift by $90^\circ$ rotates the contrast pattern accordingly [Figs.~\ref{fig:placeholder3}(c,f)], confirming that the nonreciprocal response is locked to the current direction, consistent with the axis-pinning condition $\mathbf{q}\cdot\mathbf{v}_{\mathrm{d}}=0$.

Overall, the real-space contrast $C(\mathbf{r})$ provides a direct near-field signature of drift-induced nonreciprocity, translating momentum-space symmetry breaking into spatially resolved intensity modulation. The results demonstrate that anisotropic polaritonic environments not only reshape the propagation geometry but also strongly redistribute the spatial manifestation of nonreciprocal transport, enabling directional control of near-field energy flow without structural patterning.

\section{\label{sec:conclusion}Conclusions and Outlook}

We have shown that a graphene/$\alpha$-MoO$_3$/SiC heterostructure supports electrically tunable polaritons with strongly anisotropic dispersion and controllable nonreciprocity. The hybridization between graphene plasmons and hyperbolic phonon polaritons in $\alpha$-MoO$_3$, further shaped by the SiC substrate, enables directional canalization and gate-driven topological changes in the IFCs without structural patterning.

When a dc current is applied in graphene, this anisotropic polaritonic landscape acquires a direction-dependent nonreciprocal response. Unlike drift-induced effects in isotropic environments, the presence of intrinsic in-plane hyperbolicity redistributes the nonreciprocal contrast in momentum space, so that its magnitude becomes strongly geometry-dependent while its symmetry axis remains fixed by the current direction. The resulting contrast can be systematically engineered through the drift strength, carrier density, and heterostructure geometry, providing independent control over symmetry breaking and dispersion engineering. In particular, the small-drift analysis reveals that the nodal direction is determined solely by the Doppler projection factor, whereas the amplitude is governed by a reciprocal enhancement factor associated with the underlying polaritonic environment. Finally, we note that while material losses reduce absolute polariton propagation length, the predicted nonreciprocal contrast is governed primarily by relative spectral weight redistribution and therefore remains robust under experimentally realistic damping levels.

More broadly, these results indicate that low-symmetry van der Waals heterostructures provide a route to reconfigurable nonreciprocal polaritonic media in the mid-infrared. The combination of current bias and electrostatic gating enables dynamic steering of polariton propagation, with directly observable signatures in scattering-type scanning near-field optical microscopy (s-SNOM) experiments \cite{basov2016polaritons}. Beyond the specific heterostructure considered here, the symmetry constraints and drift-induced response are expected to extend to other low-symmetry two-dimensional materials and anisotropic polaritonic platforms \cite{matson2023controlling}. Integrating natural in-plane hyperbolicity with drift-biased two-dimensional electron systems therefore offers a versatile strategy for realizing electrically controlled nonreciprocal nanophotonic functionalities, including directional antennas, anisotropic thermal emitters, and dynamically reconfigurable nanoscale waveguides.

\begin{acknowledgments}
This work has received funding from the Spanish Ministry of Science and Innovation (MICINN) through project 2D-SAWNICS (PID2020-120433GB-I00), from the Spanish Ministry of Science, Innovation, and Universities (MICIU)  through project SE2D (PID2024-159409OB-I00), and from the Comunidad de Madrid, the Recovery, Transformation and Resilience Plan of the Spanish Government and the European Union – NextGenerationEU (PRTR-C17.I1) through projects MAD2D-CM-UPM1 (Complementary Plan for Advanced Materials) and MADQuantum-CM (Complementary Plan for Quantum Communications).
\end{acknowledgments}

\bibliography{ref_list}
\end{document}